\documentclass[prx,aps,amssymb,twocolumn,superscriptaddress]{revtex4-1}
\usepackage{amsmath}
\usepackage{amssymb}
\usepackage{amsthm}
\usepackage{amsfonts}
\usepackage{listings}
\usepackage{enumerate}
\usepackage{latexsym}
\usepackage{color}
\usepackage{bm}
\usepackage{siunitx}
\usepackage{hyperref}
\hypersetup{
 pdfnewwindow=true, colorlinks=true,
 linkcolor=blue, anchorcolor=blue,
 citecolor=blue, filecolor=blue,
 menucolor=blue, urlcolor=blue}

\usepackage{psfrag}

\usepackage{bm}
\usepackage{graphicx}
\usepackage{subfigure}

\usepackage{multirow}

\RequirePackage[normalem]{ulem} 
\RequirePackage{color}\definecolor{RED}{rgb}{1,0,0}\definecolor{BLUE}{rgb}{0,0,1}

\newcommand{\webirvsp}{\href{https://github.com/zjwang11/irvsp}{\texttt{IRVSP}} }

\def\ie{{\it i.e.},\ }
\def\eg{{\it e.g.}\ }

\input{epsf}

\begin{document}
\title{Excitonic Instability in Ta$_2$Pd$_3$Te$_5$ Monolayer}

\author{Jingyu Yao}
\thanks{These authors contributed equally to this work.}
\affiliation{Beijing National Laboratory for Condensed Matter Physics and Institute of
Physics, Chinese Academy of Sciences, Beijing 100190, China}
\affiliation{University of Chinese Academy of Sciences, Beijing 100049, China}
\author{Haohao Sheng}
\thanks{These authors contributed equally to this work.}
\affiliation{Beijing National Laboratory for Condensed Matter Physics and Institute of
Physics, Chinese Academy of Sciences, Beijing 100190, China}
\affiliation{University of Chinese Academy of Sciences, Beijing 100049, China}

\author{Ruihan Zhang}
\affiliation{Beijing National Laboratory for Condensed Matter Physics and Institute of
Physics, Chinese Academy of Sciences, Beijing 100190, China}
\affiliation{University of Chinese Academy of Sciences, Beijing 100049, China}

\author{Rongtian Pang}
\affiliation{Centre for Quantum Physics, Key Laboratory of Advanced Optoelectronic Quantum Architecture and Measurement (MOE), School of Physics, Beijing Institute of Technology, Beijing 100081, China}

\author{Jin-Jian Zhou}
\email{jjzhou@bit.edu.cn}
\affiliation{Centre for Quantum Physics, Key Laboratory of Advanced Optoelectronic Quantum Architecture and Measurement (MOE), School of Physics, Beijing Institute of Technology, Beijing 100081, China}

\author{Quansheng Wu}
\affiliation{Beijing National Laboratory for Condensed Matter Physics and Institute of
Physics, Chinese Academy of Sciences, Beijing 100190, China}
\affiliation{University of Chinese Academy of Sciences, Beijing 100049, China}

\author{Hongming Weng}
\affiliation{Beijing National Laboratory for Condensed Matter Physics and Institute of
Physics, Chinese Academy of Sciences, Beijing 100190, China}
\affiliation{University of Chinese Academy of Sciences, Beijing 100049, China}

\author{Xi Dai}
\affiliation{Department of Physics, Hong Kong University of Science and Technology, Hong Kong 999077, China}

\author{Zhong Fang}
\affiliation{Beijing National Laboratory for Condensed Matter Physics and Institute of
Physics, Chinese Academy of Sciences, Beijing 100190, China}
\affiliation{University of Chinese Academy of Sciences, Beijing 100049, China}

\author{Zhijun Wang}
\email{wzj@iphy.ac.cn}
\affiliation{Beijing National Laboratory for Condensed Matter Physics and Institute of
Physics, Chinese Academy of Sciences, Beijing 100190, China}
\affiliation{University of Chinese Academy of Sciences, Beijing 100049, China}

\date{\today}
\begin{abstract}
By systematic theoretical calculations, we have revealed an excitonic insulator (EI) in the Ta$_2$Pd$_3$Te$_5$ monolayer. The bulk Ta$_2$Pd$_3$Te$_5$ is a van der Waals (vdW) layered compound, whereas the vdW layer can be obtained through exfoliation or molecular-beam epitaxy. First-principles calculations show that the monolayer is a nearly zero-gap semiconductor with the modified Becke-Johnson functional. Due to the same symmetry of the band-edge states, the two-dimensional polarization $\alpha_{2D}$ would be finite as the band gap goes to zero, allowing for an EI state in the compound. Using the first-principles many-body perturbation theory, the $GW$ plus Bethe-Salpeter equation calculation reveals that the exciton binding energy is larger than the single-particle band gap, indicating the excitonic instability. 
The computed phonon spectrum suggests that the monolayer is dynamically stable without lattice distortion. 
Our findings suggest that the Ta$_2$Pd$_3$Te$_5$ monolayer is an excitonic insulator without structural distortion.

\noindent PCAS:  71.35.–y, 71.15.Mb, 73.90.+f
\end{abstract}

\maketitle

\paragraph*{Introduction.}
The excitonic insulator (EI) is an exotic ground state of narrow-gap semiconductors and/or semimetals, arising from the spontaneous condensation of electron-hole pairs bound by attractive Coulomb interactions~\cite{Jerome1967,Shim2009,Du2017, Sun2022,Jia2022,Kong2022,Gao2023,Song2023,Wang2023,Pikulin2014,Ma2022,Shao2023}.
The excitonic instability usually happens as the excitonic binding energy ($E_b$) is larger than the single-particle band gap ($E_g$).  
Due to the Coulomb screening effect~\cite{Rohlfing1993}, the EI candidates are rare in bulk compounds.
In experiments, two kinds of bulk materials are considered as EIs, \eg, 1\textit{T}-TiSe$_2$~\cite{Cercellier2007,Monney2009} and Ta$_2$NiSe$_5$~\cite{Wakisaka2009,Wakisaka2012,Mazza2020}.
Due to the existence of the charge density wave transition or structural distortion, the origin of the phase transition in the two EI candidates is still under debate.
The plasmon softening around the transition temperature was proposed to serve as the signature of the EI in 1\textit{T}-TiSe$_2$~\cite{Kogar2017}. However, this result has not been supported by recent momentum-resolved high-resolution electron energy loss spectroscopy studies~\cite{Lin2022}. 
There is some compelling evidence for exciton condensation in artificial structures, such as InAs/GaSb quantum wells~\cite{Du2017} and MoSe$_2$/WSe$_2$ bilayers~\cite{Wang2019,Troue2023}. Specifically, the former represents the first complete experimental and theoretical confirmation of topological excitonic insulators in narrow-gap semiconductor systems. However, the experimental confirmation of the EI state in real materials remains unsolved.

On the other hand, lower dimensionality can significantly weaken the screening effect and result in a larger $E_b$. However, $E_b$ usually shows a strong dependence on $E_g$, \ie $E_b\sim E_g/4$ in two-dimensional (2D) materials~\cite{Jiang2017}. To break this dependence, one strategy is to seek dipole-forbidden transitions near the band edges~\cite{Jiang2018, Jiang2019, Jiang2020,Yang2024}.
Thus, some 2D materials are theoretically predicted to be EI candidates, such as GaAs~\cite{Jiang2018}, AlSb~\cite{Dong2021}, AsO~\cite{Yang2024}, and Mo$_2M$C$_2$F$_2$ ($M=$Ti,Zr,Hf)~\cite{Dong2023}. Interestingly, some quantum spin Hall insulators with large band inversion can result in the same-parity band-edge states. The topological EI can be achieved in such systems~\cite{Yang2024,Dong2023}. However, these 2D EI candidates still need experimental confirmation.

In this work, we demonstrate that Ta$_2$Pd$_3$Te$_5$ monolayer shows the excitonic instability by systematic theoretical calculations. 
First-principles calculations using modified Becke-Johnson functional suggest that the monolayer has a nearly zero band gap and that the band-edge states have the same $C_{2z}$ symmetry eigenvalue. Upon applying the strains, the 2D polarization $\alpha_{2D}$ shows little response to the reduction of $E_g$. 
The band gap was obtained by the $GW$ calculation with $E_g =130$ meV.
To calculate $E_b$, we performed the first-principles $GW$-BSE calculations. The obtained $E_b=633$ meV is larger than the $E_g$, indicating excitonic instability. The strain-dependent calculations show that the excitonic insulating phase is robust against small strains. 
Additionally, the tight-binding (TB) model is constructed to analyse the symmetry of the excitons.
Unlike 1\textit{T}-TiSe$_2$ and Ta$_2$NiSe$_5$, no structural instability is found in the phonon spectrum of this material. 
Our findings suggest that the Ta$_2$Pd$_3$Te$_5$ monolayer is an excitonic insulator without structural distortion.

\paragraph*{Calculation methods.}
First-principles calculations were performed within the framework of density functional theory (DFT) using the projector augmented wave (PAW) method~\cite{Blochl1994,Kresse1999}, as implemented in Vienna $\textit{ab initio}$ simulation package (VASP)~\cite{Kresse1996,Kresse1996_2}. 
20 $\times$ 4 $\times$ 1 $k$-point sampling grids were used, and the cut-off energy for plane wave expansion was 500 eV. 
Phonon spectra were obtained with the finite-difference method using a 2$\times$2$\times$1 supercell, as implemented in the Phonopy package~\cite{phonopy}.
Considering that the Perdew-Burke-Ernzerhof (PBE) exchange correlation functional~\cite{Perdew1996} underestimates the band gap, the band structures were obtained by using the modified Becke-Johnson (MBJ) functional~\cite{Tran2009,Becke2006}.
Moreover, to compute the binding energy $E_b$~\cite{Yamada2015}, first-principles many-body $GW$-BSE calculations on top of the PBE band structure were performed with the Coulomb cutoff technique in the Yambo package~\cite{Marini2009,Giannozzi2009,Rozzi2006}. 
Quasi-particle (QP) corrections in $GW$ calculations are $k$-point and band dependent.
The same $k$-point grid 
and 4 Ry cutoff were used to calculate the dielectric function matrix. 
The kinetic energy cutoff of 70 Ry was used for the evaluation of the exchange part of the self-energy. Achieving convergence of the $G_0W_0$ band gap involves employing 300 bands along with an extrapolar correction scheme~\cite{Bruneval2008}.
One valence band (VB) and two conduction bands (CBs) were included to build the BSE Hamiltonian.

\begin{figure}[!t]
	\centering
	\includegraphics[width=0.48\textwidth]{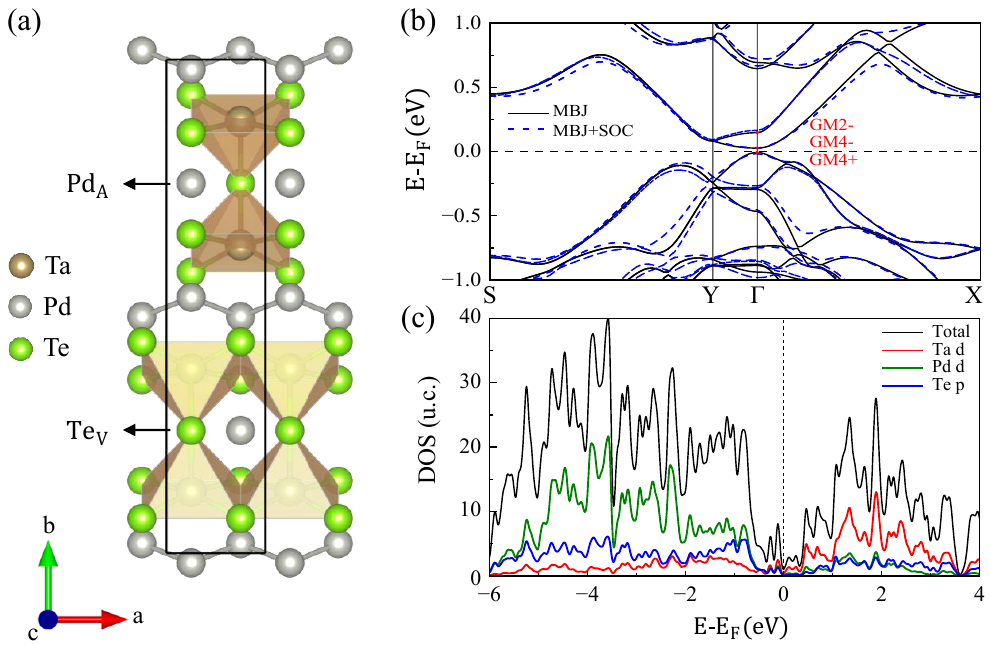}
	\caption{The crystal structure and band structures of the $\rm{Ta_{2}Pd_{3}Te_5}$ monolayer.
    (a) Crystal structure of the monolayer.
    (b) MBJ band structures with and without spin-orbit coupling. The highest VB is labeled by $v_1$, while the first and second lowest CBs are labeled by $c_1$ and $c_2$, respectively.
    (c) Total and partial DOS for Ta $d$, Pd $d$, and Te $p$ orbitals.    
    }\label{fig-pbevsgw}
\end{figure}

\begin{figure}[!t]
	\centering
	\includegraphics[width=0.48\textwidth, trim=120 0 120 0]{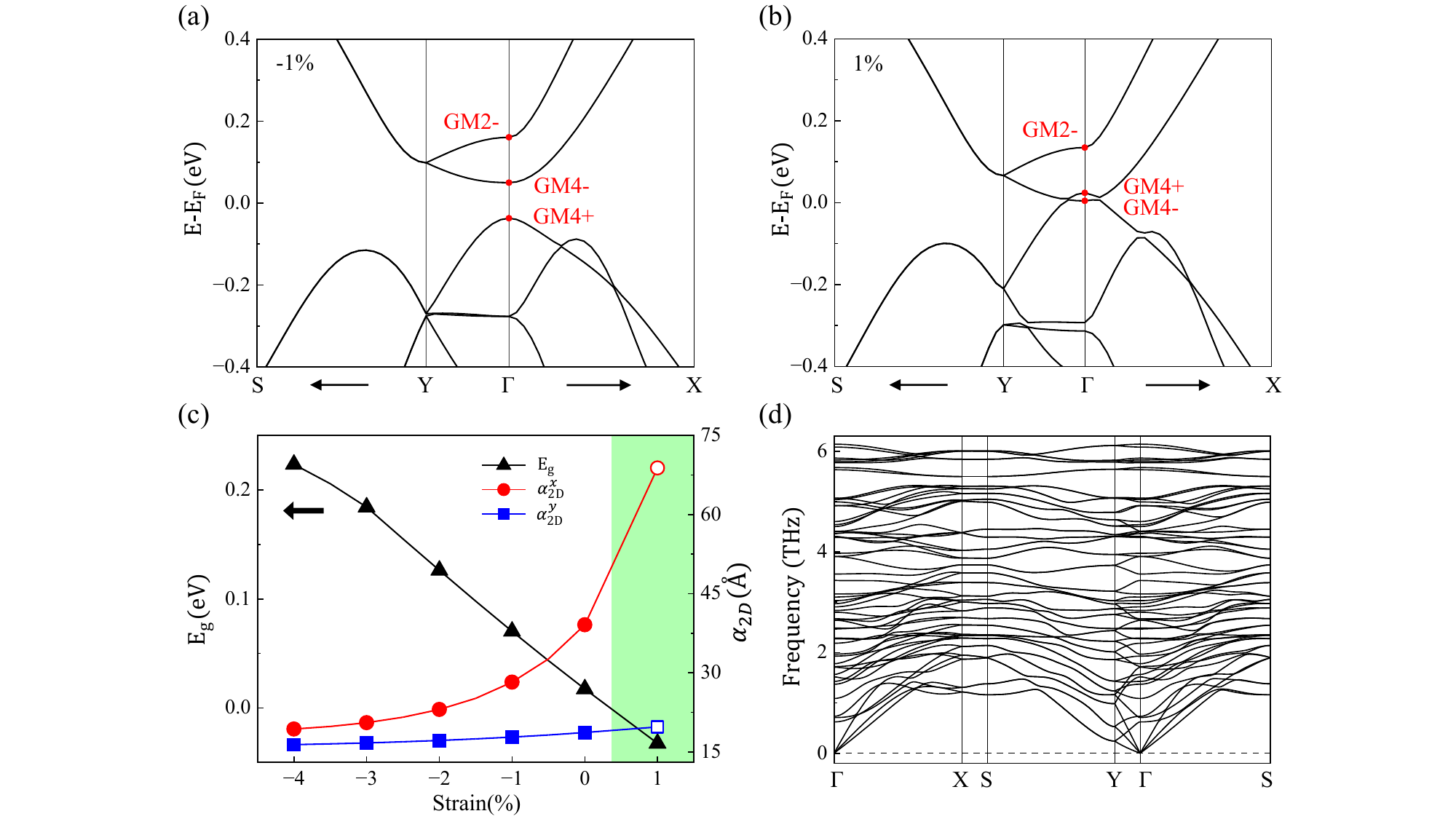}
	\caption{The evolution of band structures and polarization $\alpha_{2D}$ under uniaxial strain $\eta$, $b=(1+\eta)b_0$.
    (a,b) The band structure with uniaxial strains $\eta=-1$\% (a) and $+1$\% (b), respectively.
    (c) $E_g$ and $\alpha_{2D}$ under different uniaxial strains.
    (d) The phonon spectrum of Ta$_2$Pd$_3$Te$_5$ monolayer. There is no imaginary frequency phonon mode.
    }\label{fig2}
\end{figure}

\paragraph*{Band structure and Density of states.}
The van der Waals (vdW) layered compound Ta$_2$Pd$_3$Te$_5$ crystallizes in an orthorhombic structure with two vdW layers in a unit cell~\cite{Guo2021}. 
The monolayer can be obtained by exfoliation~\cite{Guo2021}.
The two mirror symmetries ($M_x$, $M_y$) and inversion symmetry 
are respected in the monolayer after structural relaxation.
Fig.~\ref{fig-pbevsgw}(a) shows the vdW layer structure 
with space group $Pmmn$~(\#59).
The quasi-1D chains are along the $a$ $(x)$ direction.
The phonon spectrum of the monolayer is obtained in Fig.~\ref{fig2}(d). No phonon mode with negative frequency in the phonon spectrum suggests that the monolayer is dynamically stable.

The MBJ band structure along the high-symmetry lines is presented in Fig.~\ref{fig-pbevsgw}(b). The irreducible representations (irreps) of the two band-edge states are computed as GM$4+$ ($v_1$ band) and GM$4-$ ($c_1$ band) by the \webirvsp program~\cite{Gao2021,Zhang2023}. Thus, we define the band gap at $\Gamma$ by $E_g\equiv E_{\rm GM4-}-E_{\rm GM4+}=33 $ meV, resulting in a nearly zero-gap semiconductor. 
The spin-orbit coupling (SOC) does not change the band structure at all, but slightly enlarges the band gap to 44 meV. This is because the CBs primarily originate from the Ta-$d_{z^2}$ orbitals with $J_z=0$, which have little SOC effect. Hereafter, the SOC is neglected in the following calculations.
Additionally, the symmetry eigenvalues of the three lowest energy bands are presented in Table~\ref{tablesym}. They show that the two band-edge bands both have the same $C_{2z}$ symmetry eigenvalue, although they are of different parity. Thus, the 2D polarizability $\alpha_{2D}$ can still be finite when $E_g \rightarrow 0$, breaking the strong dependence and allowing for the EI candidate with $E_b>E_g$.
Furthermore, due to the significantly  enhanced $E_b$ in lower dimensions, the Ta$_2$Pd$_3$Te$_5$ layers with the quasi-1D structure present a promising opportunity to realize an intrinsic EI. 

    \begin{table}[b!]
    \begin{ruledtabular}
    \caption{The symmetry of the highest VB ($v_1$ band) and the lowest two CBs ($c_1$ and $c_2$ bands) at $\Gamma$.}\label{tablesym}
    \begin{tabular}{cccc}
        band & irrep & $C_{2z}$ & $M_y$\\ \hline
        $v_1$&GM$4+$&$-1$&$+1$\\
        $c_1$&GM$4-$&$-1$&$-1$\\
        $c_2$&GM$2-$&$+1$&$+1$\\
    \end{tabular}
    \end{ruledtabular}
    \end{table}

The total and partial densities of states (DOS) are plotted in Fig.~\ref{fig-pbevsgw}(c). The results show that the CBs are mainly from Ta $d$ states, while the VBs are mainly from Pd $d$ states. The Te $p$ states have strong hybridization with them, and have certain contributions both below and above the Fermi level ($E_F$). In particular, the orbital-resolved band structures in Figs.~\ref{fig-exciton}(a,b) show that the CBs are contributed by Ta $d_{z^2}$ states, while the VBs are formed by the hybridization of Pd$_A$ $d_{xz}$ and Te$_V$ $p_x$ states~\cite{Guo2022}. The related bonds are $d_{\rm Ta-Pd_A}=2.99$ \AA ~and $d_{\rm Ta-Te_V}=2.82$ \AA, respectively. Although these bond hoppings are allowed, the CB and VB states do not mix on the line Y$\Gamma$ due to their different $M_x$ eigenvalues. The interaction between Ta-$d$ electrons and the Pd-$d$/Te-$p$ holes may be crucial to the formation of the EI state in the compound.

\paragraph*{Evolution of $\alpha_{2D}$ under strain.}
As we know, the band gap of this material is sensitive to the strain~\cite{Guo2021}. Figs.~\ref{fig2}(a,b) show the MBJ band structures with uniaxial strains along $y$. When the system is compressed by 1\% in Fig.~\ref{fig2}(a), the gap increases to 87 meV; in contrast, it becomes metallic under tensile strain in Fig.~\ref{fig2}(b). 
The 2D polarization, denoted as $\alpha_{2D}^{x/y}$, is calculated with the formula $\alpha_{2D}^{x/y}=c_0\frac{\varepsilon^{xx/yy} - 1}{4\pi}$, where $c_0$ is the thickness of the vacuum in the $z$ direction. The $ \varepsilon^{xx/yy}$ represents the $xx/yy$ components of the macroscopic static dielectric tensor, which is computed with the random phase approximation and considering the local field effects, as implemented in VASP.

In Fig.~\ref{fig2}(c), we plot $E_g$ and 2D polarization $\alpha_{2D}^{x/y}$ as a function of the uniaxial strain.
In the positive gap range, both show a weak dependence on the reduction of the band gap. Especially, $\alpha_{2D}^{y}$ is almost unchanged.
The weak dependence of $\alpha_{2D}^{x}$ is attributed to the transition between the $v_1$ band and the second lowest CB ($c_2$ band). The symmetry eigenvalues at $\Gamma$ yield $\left\langle c_2|\nabla_{k_y}|v_1 \right\rangle =0$ and $\left\langle c_2|\nabla_{k_x}|v_1 \right\rangle \neq 0$, which have been confirmed numerically~\cite{Zhang2023vasp2kp}.
As aforementioned, the band-edge transition between $c_1$ and $v_1$ bands is forbidden due to the twofold rotation.
This indicates the decoupling between $E_g$ and $E_b$ in this material with band-edge states of the same $C_{2z}$ symmetry.

\paragraph*{Stable phonon spectrum.}
In previous studies~\cite{Calandra2011,Bianco2015}, the phonon spectra of previous EI candidates 1\textit{T}-TiSe$_2$ and Ta$_2$NiSe$_5$ show the structural instability with imaginary frequency phonon modes. Whether the charge-density-wave transition in 1\textit{T}-TiSe$_2$ comes from the Jahn-Teller mechanism or from the excitonic instability has plagued the EI community for decades. Additionally, as indicated by the imaginary frequency mode, the structure distortion of Ta$_2$NiSe$_5$ from $Cmcm$ (SG \#63) to $C2/c$ (SG \#15) occurs at 328 K~\cite{Salvo1986,Nakano2018,Kim2016}, accompanied by a metal-to-insulator transition even in the single-particle band structure calculations. 

However, our calculation shows that there is no imaginary frequency on the phonon spectrum of Ta$_2$Pd$_3$Te$_5$ monolayer in Fig.~\ref{fig2}(d).  Even if we start from some degree of distortion, the relaxation still yields the $Pmmn$ symmetry structure. 
The phonon spectrum of the bulk Ta$_2$Pd$_3$Te$_5$ also does not show any imaginary frequency modes, quite different from the previous two examples.
Experimentally, no structural distortion is found in the X-Ray diffraction data~\cite{Guo2021}. Therefore, Ta$_2$Pd$_3$Te$_5$ inherently excludes Jahn-Teller-like instabilities, consequently avoiding the confusion of 1\textit{T}-TiSe$_2$ or Ta$_2$NiSe$_5$.
However, the lack of a structural signal usually poses challenges for identifying an EI phase transition. The symmetry breaking of the EI phase transition could be weakly coupled to the lattice structure, which needs high-resolution measurements, such as electron diffraction.

\begin{figure}[!t]
	\centering
	\includegraphics[width=0.48\textwidth, trim=10 10 0 0]{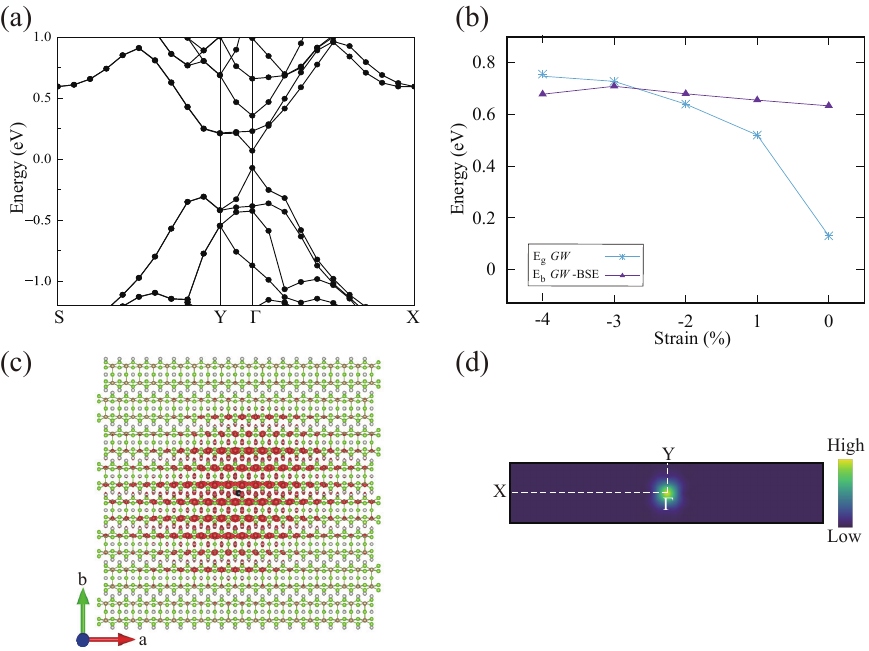}
	\caption{(a) The $G_0W_0$ band structure. 
 (b) Strain dependence of $E_g$ and $E_b$. The results show an intrinsic EI with $E_b > E_g$ at $\eta >-2$\%.
 (c) Exciton wavefunction square modulus, as obtained from the Bethe–Salpeter equation ($GW$-BSE). The contour plot (red) is the probability density of locating the bound electron once the hole position is fixed (black dot). The figure contains 20 and 4 unit cells in the $x$ and $y$ directions, respectively. We note that it is well-localized around the hole. 
 (d) Exciton wavefunction square modulus in reciprocal space. The exciton probability weight is localized around $\Gamma$ point. 
 }
	\label{fig-EbEg}
    \end{figure}

\paragraph*{Binding energy and GW-BSE calculations.}
In order to investigate the excitonic instability, we carry out many-body $GW$ calculations in a one-shot scheme ($G_0W_0$). The $GW$ band structure in Fig.~\ref{fig-EbEg}(a) does not change significantly from the MBJ one except that the $E_g$ changes from 33 meV to 130 meV. The first-principles $GW$-BSE calculation shows that the $E_b$ = 633 meV. This $E_b$ value exceeds the $GW$ band gap $E_g$. 
The lowest-energy exciton wavefunction in real space is shown in Fig.~\ref{fig-EbEg}(c) as the conditional probability of finding a bound electron (red), provided the hole position is fixed (black dot). The electron is well-localized around the hole, within a radius of 30 \AA. 
The lowest-energy exciton probability weight in momentum space is localized around the $\Gamma$ point, as shown in Fig.~\ref{fig-EbEg}(d).

In order to study the strain dependence of $E_b$, we performed the one-shot $GW$-BSE calculations under uniaxial compressive strains. 
Fig.~\ref{fig-EbEg}(b) shows $E_b$ as a function of compressive strains.
The obtained $E_b$ exceeds the $E_g$ at $\eta > -2$\%, indicating that the EI instability is robust against small strains in the Ta$_2$Pd$_3$Te$_5$ monolayer.
We observe that the obtained $E_b$ is almost unchanged, although $E_g$ varies with different strains.  
We attribute this to the unique wavefunctions of the conduction and valence states, which originate from Ta and Pd/Te atoms respectively (Fig.~\ref{fig-exciton}). 
The $E_b$ in the Ta$_2$Pd$_3$Te$_5$ monolayer shows little response to the change of $E_g$ under strain, as illustrated in Fig.~\ref{fig-EbEg}(b). 
Although the exact $E_g$ is difficult to predict and the strain condition varies in experiment, the formation of EI is accessible at modest experimental conditions.

\begin{figure}[!t]
	\centering	
    \includegraphics[width=0.48\textwidth]{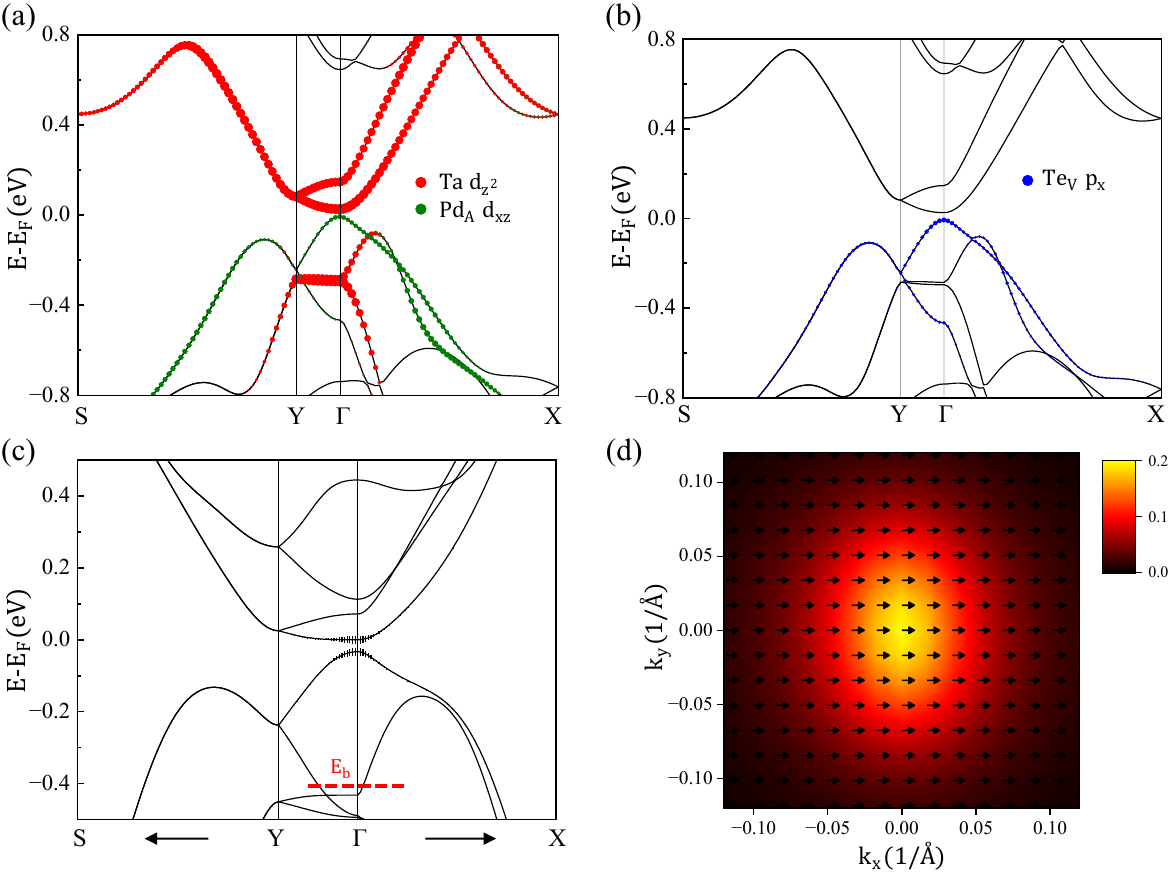}
	\caption{ (a,b) The orbital-resolved MBJ band structures for Ta $d_{z^2}$, Pd$_A$ $d_{xz}$, and Te$_V$ $p_x$ orbitals.  
    (c) The band structure of the effective TB model. The band spread corresponds to the contribution of a particular electron-hole transition to the lowest-energy exciton.
    (d) Absolute values (color) and phase angles (arrows) of the 
    lowest-energy exciton wavefunction.
    }
    \label{fig-exciton}
\end{figure}

\paragraph*{Tight-binding model and symmetry analysis.}  
In order to analyse the symmetry of the excitons, we construct a TB model and perform TB-BSE calculations to obtain the phases of the exciton wavefunctions.
In the orbital-resolved band structures presented in Figs.~\ref{fig-exciton}(a,b), we find that the valence bands are mainly formed by the Pd$_A$ $d_{xz}$ states, which hybridize with the Te $p_x$ states (especially Te$_V$). The CBs are from the Ta $d_{z^2}$ states, which do not hybridize with the valence bands along Y-$\Gamma$ line.
Accordingly, we construct a sixteen-band Wannier-based TB Hamiltonian, extracted from the DFT calculations by using Wannier90 package~\cite{Pizzi2020}. 
Under the basis of these Wannier orbitals $\{|\alpha\boldsymbol{k}\rangle\}$, the eigenvalues and eigenstates of $H_{TB}$  yield
$\hat H_{TB}(\boldsymbol{k})|b\boldsymbol{k}\rangle=E_b(\boldsymbol{k})|b\boldsymbol{k}\rangle$.
On top of the TB model in Fig.~\ref{fig-exciton}(c), we have solved the model BSE to find the collective modes. The BSE reads~\cite{Rohlfing2000,Scharf2017,Quintela2022},
\begin{equation}
\begin{aligned}
    \label{bse}
    (\Omega_S-E_c(\boldsymbol{k})+E_v(\boldsymbol{k}))
    A_{cv}^S(\boldsymbol{k})
    =
    \sum_{c^{'}v^{'}\boldsymbol{k}^{'}}
    \mathcal{K}^{cv\boldsymbol{k}}_{c^{'}v^{'}\boldsymbol{k}^{'}}
    A_{c^{'}v^{'}}^S(\boldsymbol{k}^{'}),
\end{aligned}
\end{equation}
where $c,\ v$ are the labels of  
 conduction and valence bands, 
 $\Omega_S$ is the energy of exciton eigenstates,
$\vert S \rangle \equiv 
\sum_{cv\boldsymbol{k}}A_{cv}^S(\boldsymbol{k})
\hat{c}^{\dagger}_{c\boldsymbol{k}}\hat{c}_{v\boldsymbol{k}}
\vert 0 \rangle$, and $\vert 0 \rangle$ is the non-interacting ground state.
The kernel consists of the direct part $\mathcal{K}^{d}$ and the exchange part $\mathcal{K}^{x}$
\begin{equation}
\begin{aligned}    
    \mathcal{K}^{cv\boldsymbol{k}}_{c^{'}v^{'}\boldsymbol{k}^{'}}
    &=
    {\mathcal{K}^{d}}^{cv\boldsymbol{k}}_{c^{'}v^{'}\boldsymbol{k}^{'}}
    +
    {\mathcal{K}^{x}}^{cv\boldsymbol{k}}_{c^{'}v^{'}\boldsymbol{k}^{'}},\\
    {\mathcal{K}^{x}}^{cv\boldsymbol{k}}_{c^{'}v^{'}\boldsymbol{k}^{'}}
    &=
    -V
    f_{cv}(\boldsymbol{k},\boldsymbol{k})
    f_{v^{'}c^{'}}(\boldsymbol{k}^{'},\boldsymbol{k}^{'}),\\
    {\mathcal{K}^{d}}^{cv\boldsymbol{k}}_{c^{'}v^{'}\boldsymbol{k}^{'}}
    &=
    -W(\boldsymbol{k}-\boldsymbol{k}^{'})
    f_{cc^{'}}(\boldsymbol{k},\boldsymbol{k}^{'})
    f_{v^{'}v}(\boldsymbol{k}^{'},\boldsymbol{k}).
    \\
\end{aligned}
\end{equation}
Here $V$ is the bare Coulomb potential, and $W(\boldsymbol{q})=2\pi e^2/[S|\bm{q}|(1+\alpha_{2D}|\bm{q}|)]$ is the screened Coulomb potential~\cite{Cudazzo2011,Yang2022},
where $S$ is the system area and the computed 2D polarization $\alpha_{2D}=17.854$ \AA ~is used.
We define
  $f_{b_1b_2}(\boldsymbol{k},\boldsymbol{k}^{'})
\equiv \sum_{\alpha}
\langle b_1 \boldsymbol{k}\vert \alpha\boldsymbol{k} \rangle 
\langle \alpha\boldsymbol{k}^{'}\vert b_2\boldsymbol{k}^{'}\rangle $,
$b_1,\ b_2 \in \{c,v\}$.
Since $
f_{cv}(\boldsymbol{k},\boldsymbol{k}) = 0  $
, $\mathcal{K}^{x}_{cv\boldsymbol{k},c^{'}v^{'}\boldsymbol{k}^{'}} = 0$. By solving Eq.~(\ref{bse}), one can obtain the discrete excitonic binding energies with electron-hole attractive Coulomb interactions. 
The lowest excitonic binding energy is depicted in Fig.~\ref{fig-exciton}(c), which is larger than the band gap, indicating the excitonic phase. This exciton wavefunction is given in Fig.~\ref{fig-exciton}(d), where colors and arrows indicate the distribution of absolute values and phase angles, respectively.  It is $1s$-like in the vicinity of the $\Gamma$ point, which is consistent with the DFT result in Fig.~\ref{fig-EbEg}(d). As the band-edge states belong to $\rm GM4+ $ and $\rm GM4- $ irreps respectively, the $1s$-like excitonic state breaks spatial inversion and all mirror symmetries.

\paragraph*{Discussion.}
In this work, we demonstrate that the Ta$_2$Pd$_3$Te$_5$ monolayer is an excitonic insulator by first-principles $GW$-BSE calculations. In the single-particle picture, the MBJ calculation shows that the monolayer is a nearly zero-gap semiconductor. The low-energy states at $\Gamma$ have the same $C_{2z}$ symmetry eigenvalue, making the band-edge transitions forbidden and keeping the $\alpha_{2D}$ finite as $E_g\rightarrow 0$. By applying the uniaxial strains, the $\alpha_{2D}$ shows little response to the reduction of $E_g$ as expected.
The $G_0W_0$ band gap $E_g=130$ meV and $E_b=633$ meV is obtained by performing one-shot $GW$-BSE calculations in Yambo, indicating an intrinsic EI with $E_b > E_g$. 
By investigating the strain effect, we find that the strong excitonic instability is robust against small strains, and the $E_b$ shows little response to the change of $E_g$.
Although the $E_b$ may be slightly modified in the self-consistent $GW$-BSE calculations, the EI phase would form as the gap goes to zero under strain.
Therefore, we conclude that the excitonic insulator phase in the Ta$_2$Pd$_3$Te$_5$ monolayer is achievable in experiments.

In the vdW layered Ta$_2$Pd$_3$Te$_5$ bulk, the band-edge states at $\Gamma$ have the same parity in the bulk MBJ band structure, making transitions between them forbidden. The decoupled relationship between $E_g$ and $E_b$ remains, making the EI phase possible in the bulk crystals.
In addition, the layered compound possesses several advantages.
To begin with, in the series of the $A_2M_3X_5$ ($A$ = Ta, Nb; $M$ = Pd, Ni; $X$ = Se, Te) family, the band gap is modified by chemical doping. Unlike the small gap semiconductor Ta$_2$Ni$_3$Te$_5$ with higher-order topology~\cite{Guo2022} and the metallic compound Nb$_2$Pd$_3$Te$_5$ with superconductivity~\cite{Higashihara2021}, the Ta$_2$Pd$_3$Te$_5$ is a nearly zero-gap
semiconductor, exhibiting strong EI instability.
Furthermore, it is a vdW layered compound with 1D chains and strong anisotropy, where the screening effect is relatively weak, resulting in a large excitonic binding energy. 
Moreover, the chemical potential of the crystals is right in the tiny band gap in experiments, showing the ideal balance of electrons and holes for excitonic condensation. 
Finally, the layered compound is easy to exfoliate and to fabricate into devices, and its properties can be readily tuned by gate voltage.  Therefore, we conjecture that the EI state is also promising in the few-layer flakes and bulk samples.

\paragraph*{Acknowledgements.}
We thank Xinzheng Li, Huaiyuan Yang, and  Yuelin Shao for helpful discussions. This work was supported by the National Natural Science Foundation of China (Grants No. 11974395, No. 12188101), the Strategic Priority Research Program of Chinese Academy of Sciences (Grant No. XDB33000000), National Key R\&D Program of China (Grants No. 2022YFA1403800 and No. 2022YFA1403400), and the Center for Materials Genome.

\paragraph*{Note added.}
In the preparation of this work, two collaborative experimental studies have been conducted on the bulk crystals~\cite{Huang2024,Zhang2024excitonic}.  At the finalizing stage, we noticed another related experimental work~\cite{Hossain2023}. 
%
        
\end{document}